\documentclass[a4paper,11pt]{article}

\usepackage{jheppub} 

\usepackage[T1]{fontenc} 

\title{A spacetime discretization and black holes properties  in a holographic representation.}

\author[a,1]{D. Pigato\note{Corresponding author.}}
 
\affiliation[a]{UBA, I-13897 Occhieppo Inferiore (Bi), Italy.}

\emailAdd{daniele.pigato@polito.it}

\abstract{We investigate the black holes properties with a very simple and semi-classical model of spacetime discretization.
In this context, we apply the Heisenberg's uncertainty principle and the equipartition energy theorem to thereto, obtaining the same thermodynamical relation of general relativity, without resorting to the Einstein field equations.\\
This fact, show us a possible clue of convergence between general relativity and quantum physics,  which we believe becomes manifest in this class of compact objects.
\\
Finally, we apply the holographic principle to thereto, introducing the concept of surface information density and advancing the idea that black holes can be considered as very simple objects, which maximize the energy and the information content for unit of spacetime.
}

\begin{document}
\maketitle
\flushbottom

\section{Introduction and Basic Assumptions}
\label{sec:intro}

Black holes, spacetime and quantum gravity, remain some of the most important and unresolved problems of modern physics.\\
Although, in the last decades, many studies and theories have been made in order to unify general relativity and quantum physics, such as, for example, the superstring theory and the loop quantum gravity, many aspects still remain debated and not yet well defined.
\\
This is the reason for which we approach this problem from a new point of view. In particular, we present a very simple model of spacetime discretization, in term of Planck units, where we apply the Heisenberg's uncertainty principle and the equipartition energy theorem, in order to determine the spacetime energy content.
\\
The analysis is performed by studying black holes properties from a semi-classical point of view. This is because we believe they can be considered as ideal systems, where the intimate nature of the spacetime becomes manifest, due to their extreme compactness and gravitational field.

Black holes can be ideally divided in three big categories, depending on the mass of the system: stellar-mass black holes ($\approx 3 \div 20 M_{\odot}$), intermediate black holes ($\approx 10^2 \div 10^6 M_{\odot}$) and giant black holes ($M> 10^6 {\odot}$), inside the nucleus of many  galaxies.
\\
Nothing about the internal structure and the distribution of the energy density is available from the outside, even if some recent works \cite{rovelli,rovelli2} suggest the idea that black holes can be considered as an intermediate state between the gravitational collapse and a bounce phase, achieved after which the energy density has reached the planck density inside the compact object.
However, due to their extreme gravitational red-shift, this process appears extremely long for an external observer and of the order of magnitude predicted by Hawking.

We start this analysis form a relativistic geometrical description of spacetime, following the well known Einstein field-equations:
\begin{eqnarray}
G_{\mu \nu}  \equiv  R_{\mu \nu} -\frac{1}{2}Rg_{\mu \nu}= \frac{8 \pi G}{c^4}T_{\mu \nu}  \label{eq:TRG} \, ,
\end{eqnarray}
where $R_{\mu \nu}$ is the Ricci tensor,  $g_{\mu \nu}$ is the metric tensor, $G$ is the Newton's gravitational constant, $c$
is the speed of light in vacuum, $R$ is the scalar curvature and $T_{\mu \nu}$ is the stress–energy tensor.
\\
In this framework, we consider the most simple case of a spherical, non-rotating and not charged mass, enclosed by an imaginary surface which is the
event horizon (EO) of the black hole \cite{nohair,hawking1}.

In the first part of this work, we suppose all the physical information about the original system lost during the gravitational collapse and the total energy converted in the mass-energy of the BH, in agreement with the equation: $E=Mc^2$. \\
Contrariwise, in the last Section, we will explore the possibility that such information might  not be lost in a singularity, but stored in the EO of the BH as bits of information in a holographic description of the system and eventually re-emitted during the evaporation process \cite{evap1,evap2}
\\
According to Einstein's theory of general relativity, spacetime has no intrinsic properties other than its curved geometry. In particular, in spherical symmetry, the line element of the metric is well described by the Schwarzschild solution:
\begin{eqnarray}
c^2 d\tau^2 =
(1- \frac{r_s}{r})c^2dt^2 -(1-\frac{r_s}{r})^{-1} dr^2 -r^2 d\Omega^2 \label{eq:Swardshild_metric} \, ,
\end{eqnarray}
where $\tau$ is the proper time, $t$ the time measured by an observed at infinity, $r$ the radial coordinates, $d\Omega^2= (d\phi^2 + sin(\phi)^2 d\varphi^2)$ with $\phi$ and $\varphi$ the colatitude and longitude angle respectively, and $R_S=2GM/c^2$ the Schwarzschild radius of the massive body.
Any non-rotating and non-charged mass with a radius equal to $R_S$, forms a black hole.
\\
In this framework, although in general relativity the concept of surface gravity is not well defined, for a stationary black hole this is not true and it is always possible associate a killing horizon to the black hole  \cite{wald}.
Therefore, we can always choose a null hyper-surface, defined by the vanishing of the norm of a Killing vector field on the event horizon, where we can calculate it.
In this condition, the surface gravity $g$ on a killing horizon, is the acceleration as exerted at infinity needed to keep an object at the horizon. Mathematically, if $k^{\alpha}$ is a suitably normalized Killing vector, the surface gravity is defined as:
\begin{eqnarray}
k^{\alpha} \nabla_{\alpha} k^{\beta} = gk^{\beta}  \label{eq:killinghorizon} \, ,
\end{eqnarray}
where the equation is evaluated at the killing horizon. \\ For the Schwarzschild solution, we take $k^{\alpha}$ to be the time
translation Killing vector $k^{\alpha} \partial_{\alpha}= \partial/\partial t$.
\\
In particular, in the limit of an adiabatically slow evolution, we reobtain  the usual thermodynamical law giving the local increase of the entropy of a fluid element heated by the dissipations associated to viscosity and the Joule's law \cite{joullaw}.
In these conditions, the surface gravity takes the simple form of $g=1/4M$ or in SI $g=c^4/4GM$, which coincides exactly with the newtonian expression  \cite{jacobs,carter,hawkingg,raine}. The Schwarzschild radius ($R_S = 2GM/c^2$), can instead be written in a more compact and useful form, in terms of  linear density $K_u = c^2/2G =M/R_S$, which represents a universal property of any Schwarzschild BH.
\\
The event horizon constitutes therefore an equipotential surface, where we define a gravitational potential $\Phi_{(EO)}= -GM/R_S$, constant for any Schwarzschild BH and proportional to the square of the escape velocity ($\Phi_{(EO)}= -GK_u \,=\, -c^2/2$). Furthermore, it is always possible to associate to any equipotential surface a gradient operator ($\nabla \Phi(d) \,=\, GM/d^2$) which corresponds to the local gravitational field. In particular, on the event horizon, we can identify an equivalent newtonian acceleration given by $\nabla \Phi_{(EO)} \,=\, GK_u/R_S$ and we can calculate the corresponding laplacian operator as $\nabla^2 \Phi_{(EO)}= 2GM/R^3_S \,=\, 1/t^2$, which represents the light extension of the BH.
\\
With this formalism, the average density and energy density can be written as: $<\rho>= 3K_u/4 \pi R^2_S$ and
$<\epsilon>= 3K_uc^2/4\pi R^2_S$. Naturally, these expressions represent only the average of these quantities and do not consider
the real matter-energy distribution inside it (which do not constitute the main goal of this work).
Nevertheless, from an external point of view, this does not influence the general properties of the BH, but allow us to highlight their dependence from the inverse of the area of the EO and not from its volume.
As we will show later, this fact opens important consequences in their physics, showing possible analogies with an holographic description of the system. \cite{holo1,holo2,verlinde1,verlinde2,HoloHooft,string}.

\section{A spacetime discretization}
\label{sec:discretization}

Let's introduce a discretization of the spacetime in terms of Planck units. Namely, we consider a spacetime characterized by
minimum dimensions: a Planck length ($l_p = \sqrt{\hbar G/c^3} \approx 1.61 \cdot 10^{-35}$ m), a Planck mass ($m_p=\sqrt{\hbar c/G} \approx 2.18 \cdot 10^{-8}$ Kg) and a Planck time ($t_p =l_p/c= \sqrt{\hbar G/c^5} \approx 5.39 \cdot 10^{-44}$ sec). These quantities, given by the elementary constants $G$, $c$, $\hbar$, constitute a natural basis from which construct the Planck metric (the spacetime metric).
\\
Following this discretization, we can easily construct a semi-classical model of a BH, which is mathematically consistent with the results obtained through general relativity.

At this regards, let us assume the fundamental unit of spacetime equal to $l_p$ ($t_p$). In this way, it is easy to show that a minimum radius must exist at which a massive body (or energy) can be confined in order to forms a BH. In fact, being $R_S=2GM/c^2$, we get a sphere of a minimum radius of $R^{min}_S= 2R_g =2l_p$, where a Planck mass-energy can be enclosed (with $R_g=c^2/G$ the gravitational radius).
\\
In this context, we can image a black hole as made up by $n$ elementary units, or Planck black holes, where $n=R_S/2l_p$ or $n=M/m_p$, are the linear degrees of freedom of the system.
In particular, in order to not violate the linear relation $K_u=M/R_S$, the Schwarzschild radius can be thought as composed by $n$ elementary units, each of which can be ideally parameterized as containing an equivalent energy of $E_p/2$ (with $E_p=m_pc^2$ the Planck energy), which saturates the energy of spacetime at the Planck level.
\\
Naturally, this represents only a mathematical discretization of spacetime, useful to underline and simplify some physical properties of the system and not its quantization.

We can show this in more detail and in a quantitative way by resorting to the Heisenberg's uncertainty principle: $\Delta E \Delta t \geq \hbar/2$. \\ In fact, being the maximum theoretical energy per unit of time given by $E = \hbar/2t$, writing it in term of Planck unit ($t_p=l_p/c$) and remembering that $l_p=R^{min}_S/2 =Gm_p/c^2$, we get:
\begin{eqnarray}
E = n \frac{\hbar c^3}{2Gm_p} \,\, =\,\, n\frac{E_p}{2} \label{eq:indet} \, ,
\end{eqnarray}
which represents the maximum possible energy for unit of spacetime ($l_p$ or $t_p$) in the Schwarzschild metric.
\\
This represents a very important result and allows us to show that the Schwarzschild metric is a propriety of spacetime, which emerges from the Heisenberg's uncertainty principle in the Planck metric.
\\
In fact, as already observed, the maximum possible energy can not exceed a value of $E=E_p/2$ for unit of Planck.
Therefore, a Planck energy ($E_p=m_pc^2$) will be contained in a sphere of minimum radius of $R=2l_p$, which is exactly the Schwarzschild radius of a Planck BH ($R^S_{min}= 2\hbar/m_pc \,=\, m_p/K_u$).\\
In this sense, the Schwarzschild metric emerges from the discretization of the spacetime in Planck units and the Heisenberg's uncertainty principle.
This fact is very important and somewhat surprising and can constitute a possible clue of convergence between  general relativity and quantum physics.

In this framework, the Heisenberg's uncertainty principle, gives us information not only about the content of energy per unit of spacetime (for $t\rightarrow t_p$ the quantum fluctuations are of the order of magnitude of $E_p$), but it can also be considered as the source of the vacuum energy.
\\
In this sense, we can image the empty space as full of this "virtual-potential" energy, which is normally negligible or close to be, except when the scalar curvature becomes significative. At this point, quantum fluctuations do not annihilate each other completely and provide a non-zero contribution to the average energy density of the outer space (see next Section).
\\
In this scheme, black holes can be thought as systems which saturate the energy of spacetime at the maximum quantity possible for unit of Planck, in agreement with the Heisenberg's uncertainty principle.  Hence, the spacetime enclosed by the event horizon is degenerated and constitutes the maximum configuration of energy possible. Naturally, even the outer spacetime possesses an high energy per unit of Planck, but it does not saturate at the Planck level.
\\
In this context, the macroscopic properties of these compact objects are determined, at microscopic level, by the discretization of the black hole in term of Planck units. Therefore, the total mass and the radius of a black hole, can be written as: $M_{BH}=nm_p$ and $R_S=2nl_p$, where, for $n=1$, we get the elementary unit of black hole, namely the Planck black hole.

\section{Black holes proprieties}
\label{sec:thermodynvar}

Following the above prescriptions, we are now able to introduce important thermodynamical variables, such as for example the temperature and the entropy of the system, calculated on its killing horizon.
\\
At this purpose, let us remember that the event horizon is considered here as an ideal, isothermal and static surface, subject to quantum fluctuations, which emits as a perfect black body ($\epsilon=1$) in a quasi thermodynamical equilibrium \cite{hawking1,hawking2,Lbh}.
\\
The emitted power is supposed to be equally divided among the $N$ cell elements which constitute the event horizon, where $N$ are the surface degrees of freedom of the system and are related to the linear degrees of freedom $n$, by the simple relation: $N=16 \pi n^2$, with $N=Ac^3/G \hbar \,=\, A/l^2_p$. Let us observe that, when the system has radiated an equivalent amount of energy equal to $E_p$, the event horizon has been reduced of one unit of Planck area ($A_1= 16 \pi l^2_p$), which corresponds, in ours parametrization, to a reduction of $2l_p$ in the Schwarzschild radius.
\\
Under these assumptions, we can calculate the temperature at which the BH emits by resorting to the equipartition energy theorem.
\\
Being $E=Nk_bT/2$, with $N=16 \pi n^2$ and $E=nm_pc^2$, we get an emission temperature of $T=E_p/(8\pi n k_b)$, which returns:
\begin{eqnarray}
T_{BH}  = \frac{1}{8\pi n}T_p  \,\,    \label{eq:TBHridotta} \, ,
\end{eqnarray}
where $n=M_{BH}/m_p$ and $T_p=\sqrt{c^5\hbar/G}/k_b\,=\, m_pc^2/k_b$ is the Planck temperature.

At this point, from the Stefan-Boltzmann equation applied to a perfect black body ($\epsilon=1$), in the aforementioned hypothesis of spherical symmetry and isotropic emission, we can calculate the luminosity as follow: $L= \epsilon A \sigma T_{BH}^4$, where $T_{BH}=T_p/8\pi n$ is the temperature, $A=16 \pi n^2 l^2_p= Nl_p^2$ the area of the event horizon and $\sigma=(\pi^2k_b^4/60 \hbar^3c^2)$ the Stefan-Boltzmann constant.
\\
At this point, it is straightforward to obtain: $L_{BH}=2K_uc^3/(15360 \pi n^2)$. For simplicity we rewrite this expression as follow: being $L_0 = \hbar c^6/2G^2m^2_p =  K_uc^3 \,=\, 1.82 \cdot 10^{52}$ W, we get $L_{BH}=\hbar c^6/(15360 \pi G^2M^2)$, with $M^2=n^2m_p^2$, which is the same expression obtained through general relativity \cite{Lbh}.
\\
We can rewrite it in a more compact and useful form,  as follows:
\begin{eqnarray}
L_{BH}  = \frac{L_0}{7680 \pi n^2}  \,=\, \frac{L_0}{480 \, N}   \label{eq:Lbh} \, .
\end{eqnarray}
\\
Therefore, a black hole emits at extremely low power of $L=L_0/480N$ and at a temperature of $T=T_p/8\pi n$.
\\
Let us observe the dependence of $T_{BH}$ from the inverse of the linear degrees of freedom of the system ($T_{BH} \propto 1/n$) and of the luminosity from the inverse of the square of the linear degrees of freedom $L_{BH}\propto 1/n^2$.\\
In this context, the temperature follows the same simple linear relation which characterizes its mass-radius dependence ($M/m_p \,=\, R_S/2l_p \propto n)$, whereas the luminosity, like the average density and the energy density, depends on the surface degrees of freedom $N$.
As we will see in next Section, this behavior can be well described in terms of holographic principles.
\\
Naturally, the evaporation process is continuous and not discrete. Therefore, even if we have considered the black hole as composed by $n$ elements, each of them parameterized as containing an half of the Planck energy, it emits with a continuous emission spectrum.

At this point, it is interesting to compare Eq. (\ref{eq:TBHridotta}) with the thermal emission predicted by Hawking \cite{hawking1,hawking2} and by Unruh \cite{unruh1,unruh2,donoghue}, following two phenomenological different approaches.\\
As know, Hawking's radiation represents the thermal emission predicted to be released near the event horizon ($T_H=\hbar c^3/8\pi GMk_b$), whereas Unruh's one, constitutes the thermal emission measured by a non-inertial observer in the vacuum ($T_U=\hbar g/2\pi c k_b$).
\\
Although phenomenologically different, these two expressions converge on the event horizon and can be expressed as:
\begin{eqnarray}
&&T_{H} \equiv T_U =  \frac{1}{M_{BH}} \frac{\hbar c^3}{8\pi G k_b}    \,\, \label{eq:TH}  \, .
\end{eqnarray}
%
Naturally, in the limit of $n=1$ ($M_{BH}=n m_p$), Eq. (\ref{eq:TH}) reduces to $T_H= T_p / 8\pi$, which corresponds exactly to Eq. (\ref{eq:TBHridotta}). \\
It is somewhat surprising to be able to obtain the same results of Hawking and Unruh, with this very simple model, only resorting to the Heisenberg's uncertainty principle and the equipartition theorem in order to determine the energy spectrum of the black hole in the Planck metric.
\\
In agreement with these results, we redefine the Unruh temperature by requiring its reduction to Eq. (\ref{eq:TBHridotta}) on the EO.
Multiplying and dividing $T_U$ by $c^2$ and remembering that $G\hbar/c^3=l^2_p$ and $m_pc^2/k_b=T_p$, we get $T_U= n/d^2 \cdot ( T_pl^2_p/2\pi)$. Now, being $T_p=8\pi n T_{BH}$, the modified Unruh temperature can be written as:
\begin{eqnarray}
&&T_{g} =   T_{BH} \frac{R^2_S}{d^2}  \,\, \label{eq:Tg}  \, .
\end{eqnarray}
This temperature, just like the Unruh's one, gives us a measure of the quantum fluctuations at which the vacuum is subjected, in presence of an external  gravitational field.
Naturally, when $d=R_S$, it reduces to Eq. (\ref{eq:TBHridotta}), in agreement with the previous results.
\\
So, Eq. (\ref{eq:Tg}) is more general than Eq. (\ref{eq:TBHridotta}) and allows us to estimate the thermal radiation associated to a gravitational field in the empty space at any distance $d$ from it.
More precisely, it provides a measure of the energy present per unit of spacetime in the outer space of the event horizon.\\
In this context, $T_g$ can be considered as a measure of the vacuum energy and therefore of the spacetime curvature ($R=g^{ij}R_{ij}$) of that region of spacetime.
\\
In particular, quantum fluctuations, induced by the Heisenberg's uncertainty principle in the scalar curvature, become manifest as Unruh's temperature and represent a measure of the vacuum energy of that region of spacetime.
\\
In this sense, the only difference between the spacetime enclosed by the event horizon and the outer one, is that, outside, the energy per unit of spacetime does not saturate at the Planck level.
In any case, we can always associate an equivalent temperature given by Eq. (\ref{eq:Tg}) to it and hence an energy of $E_{T}=k_bT_g$, which corresponds to an equivalent mass for unit of length of $m_{T}=E_{T}/c^2$.  Naturally, the ratio $m_{T}/2l_p$ will be lower than $K_u$, for example at $d=2R_S$ from the EO of a generic BH we get: $m_{T}=m_p/8$ (in fact spacetime is not degenerated in energy).
\\
So, the outer space, although not degenerated in energy, posses an high equivalent energy per unit of length. This energy becomes manifest as the Unruh's radiation and finds its source in the quantum fluctuations generated by the Heisenberg's uncertainty principle in the spacetime curvature.
In this framework, spacetime can be considered as a "dynamical medium of energy storage". In particular, when the curvature is zero, quantum fluctuations cancel out each other statistically, giving a zero contribution to the local average energy density. Whereas, when the energy density increases up to achieves the Planck level, the curvature becomes large enough to form a black hole and trap this energy inside the event horizon.
\\
This fact, allows us to clarify the reason for which, the average energy density of the Universe appears extremely small if compared to what predicted by quantum considerations. Experimentally, its value corresponds to about $\epsilon \approx 10^{-9}$ J/m$^3$, namely $100$ orders of magnitude lower than what expected by the standard cosmological model. This fact, supported by WMAP and PLANK data \cite{WMAP,PLANK}, is in agreement with this model and suggests a very small or negligible value of scalar curvature for the Universe and hence a very small value of the vacuum energy density, which reflects to a critical density close to zero ($\Omega_{crit} \approx 1$).

In the following, we will use Eq.s (\ref{eq:Lbh}) and (\ref{eq:Tg}) to obtain the evaporation lifetime and the entropy of the BH.
\\
At this regard, let us remember that the total power radiated by the black hole is very small and decreases with the square of the linear degrees of freedom of the system, in agreement with Eq. (\ref{eq:Lbh}).
Therefore, the lifetime of the black hole can be simply calculated through the well know relation: $L=-dE/dt= -c^2 dM/dt$, with $L=L_0/480N$. Making explicit the mass of the compact object $L=\hbar c^6/(15360 \pi G^2M^2)$ and integrating over $dM$ and $dt$, we get:
\begin{eqnarray}
t_{ev}= \frac{5120 \pi G^2M^3}{\hbar c^4}   \,=\, 5120\pi \cdot n^3 t_p   \,\, \label{eq:tvita}  \, ,
\end{eqnarray}
with $G^2m_p^3/ \hbar c^4 \,=\, m_p/ 2K_u c \,=\, t_p$.
\\
Note that this time grows very fast, with the cube of the linear degrees of freedom of the system ($t_{ev} \propto n^3$). In this context, a Planck black hole ($n=1$) has an expected lifetime of $t_{ev}=5120\pi t_p \approx 10^{-39}$ sec and an emission temperature of $T=T_p/8\pi \approx 5.6 \cdot 10^{30}$ K, whereas, a $n=10$ black hole,  presents a life time three orders of magnitude bigger and a corresponding emission temperature one order of magnitude lower.
\\
Following this result we observe that, for macroscopic objects, the evaporating life time becomes extraordinary long (for example a solar mass black hole is expected to have a life time of about $t_{ev} \approx 10^{74}$ sec).
\\
These results are very important and clarify the impossibility to observe this evaporation process for stellar mass black holes.
However, in presence of hypothetical mini black holes, generated in the first stages of evolution of the Universe, it can be observable right now how as an excess of gamma radiation in the background \cite{hawking1,minibh1,minibh2,minibh3}.
In fact, for a black hole of mass $M \approx 200 \cdot 10^{9}$ Kg, we obtain a lifetime of about $t \approx 13.7 \cdot 10^{9}$ years, which is approximately the age of the Universe.

We are now able to calculate the Entropy of the system. \\
Again, suppose the event horizon as an ideal and isothermal surface, subjected to quantum fluctuation in agreement with the Heisenberg's uncertainty principle. Near to thermodynamic equilibrium, the variation of  entropy $S$ of the system is related to the energy $E$ and the work done by the external agents $dW$ by the first law of thermodynamics: $TdS=dE-dW$. However, since we study a not charged and not rotating BH, the term $dW=\Omega dJ + \Phi dQ$ is zero by definition, where $\Omega dJ$ and $\Phi dQ$ are respectively the rotational and the coulomb energy of the system \cite{entropy1,entropy2,entropy3,entropy4,omnisbh}.
Therefore, the entropic variation in the microstates of the system, only depends on the variation in the black hole energy spectrum.
\\
In this condition, the first principle of thermodynamics reduces to $T dS=c^2 dM$, with $T=T_p/8\pi n$ and $n=M/m_p$. 
Now, substituting and integrating over $dM$ and $dS$, we get:
\begin{eqnarray}
&&S_{BH}=  \pi \frac{k_b c^3}{G \hbar} R_S^2 \,=\, \pi \frac{k_b}{l_p^2} R_S^2 \,=\, 4\pi n^2k_b   \,\, \label{eq:Sbh}  \, .
\end{eqnarray}
Naturally, in the limit of  $n=1$ ($R_S=2n l_p$) we obtain $S_{1}=4\pi k_b$, which represents the entropy of a Planck BH.
Therefore, due to the equipartition of the energy and in agreement with the spacetime discretization adopted and the Heisenberg's uncertainty principle, the total entropy of the BH can be expressed as $S_{BH}=4\pi n^2 k_b \,=\, Nk_b/4 $.
\\
This represents a very important result and it is also consistent with the Jacob-Bekenstein inequality (JBi) $S_{JB} \leq (2\pi k_b/\hbar c)\cdot R_S E_{BH}$, which represents an upper limit to the entropy $S$ that can be contained within a given finite region of spacetime \cite{bekenstein1}. In particular, in the limit of $R=R^{S}_{min}$ and $E=E_p$, we get an upper limit of $S=(\pi k_bc^3/ \hbar G) R^2_S$, which gives $S=4\pi k_b$, when $n=1$.
\\
This expression, corresponds exactly to  Eq. (\ref{eq:Sbh}) and to the Bekenstein-Hawking entropy $S_H= (k_b c^3/4G \hbar) \cdot \int_S dA$ \cite{hawking1,entropy2}, obtained in presence of non-rotating and not-charged black hole, which saturates the bound.\\
This result constitutes another important factor in support of the idea that black holes can be considered as very simple objects, which can be parameterized as constituted by $n$ elementary Planck units, or Planck black holes.
\\
In this context, their macroscopic proprieties, such as for example the mass, the radius, the temperature and the energy, can be expressed in terms of linear degrees of freedom $n$, whereas the entropy, the energy density and the luminosity by $n^2$ ($N$). The lifetime is instead proportional to $n^3$.

\section{Holographic elements}
\label{sec:holog}

In the previous sections we have used the energy equipartition theorem and  the Heisenberg's uncertainty principle, in order to obtain a coherent description of the macroscopic properties of black holes in the Planck metric.
\\
In this section, we would like to investigate some properties of black holes and therefore of spacetime, within an holographic description of the system \cite{holo1,holo2,verlinde1,verlinde2,HoloHooft,string}.
In this sense, some possible indications have already been shown in the previous sections, in particular for what concerns the dependence of the average energy density and of the entropy of the system from the inverse of the area of the EO, rather than its volume ($<\epsilon> \, \propto  \epsilon_p/N$ and $ S \propto N S_p$).
\\
This behavior constitutes a possible manifestation of the holographic properties of spacetime and allows us to express many physical parameters through an holographic description of the system.

At this regards, let us remark some important properties of holography.\\
The holographic principle states that: everything that is embedded in a space region can be described by bits of information on its border \cite{holo1,holo2,HoloHooft,string}.  Therefore, studying the physics on the holographic surface, corresponds to studying the physics of its volume. Furthermore, the total amount of information stored on the holographic surface cannot exceed a maximum limit, which corresponds to the total number of bits of information storable on the surface of the event horizon. In this context, through the spacetime discretization adopted here and the Heisenberg's uncertainty principle, the maximum number of information possible for unit of Planck can not exceed a corresponding energy of $E=E_p/2$.
This constitutes a very important result. In fact, in analogy with the Einstein equivalence principle ($E=mc^2$), it highlights an equivalence between the energy and the information stored in the system. In particular, through the first law of thermodynamics we get the thermodynamical relation: $E= 2ST$, with $S=4\pi n^2k_b \,=\,$ $|I|$, where $I$ is the Shannon information present in each element of the event horizon.
Therefore, add a bits to the system, corresponds to an increase of the surface of the event horizon of one unit of Planck area, which corresponds to $l^2_p$. In this scheme, the total energy of the system can be expressed in terms of linear or surface degrees of freedom as $E=n k_bT_p$ or $E=N \cdot k_bT_{BH}/2$, which corresponds to the equipartition of the energy among the N units which constitute the EO.
\\
In this context, the total amount of information storable in a system is given by $S=4 \pi n^2k_b \,= \, N k_b/4$, with $S=|I|$. Therefore, a Planck black hole ($n=1$), constitutes the minimum value of entropy and therefore information storable in a degenerated spacetime.
\\
Furthermore, let's note that, to an increase of the entropy corresponds an equivalent reduction of the information achievable in the system and vice versa ($S=-I$).
However, in a holographic representation, such information might not be lost during the gravitational collapse, but it could be conserved in the holographic screen (the event horizon) of the black hole and eventually released in terms of elementary particles and radiation during the evaporation process  \cite{evap1,evap2}.

In this framework, we introduce a new and important physical parameter which characterizes the information stored in the black hole and that we call surface information density ($\rho_I$).
\\
It represents the ratio between the total entropy and the area of the event horizon. 
Being $|I|=S \,=\,4\pi n^2k_b$ and the area of the event horizon $A=16\pi n^2l_p^2$, we get:
\begin{eqnarray}
&& \rho^{EO}_I   \,=\, \frac{S}{A}=\frac{1}{4} \frac{k_b}{l_p^2} \cong 1.3 \cdot 10^{46} (\frac{J}{m^2 \cdot K})    \,\, \label{eq:SA}  \, .
\end{eqnarray}
Note that this quantity is a fundamental constant for any Schwarzschild BH. In fact, it is independent from the degrees of freedom of the system (this fact underlines the utility of expressing the physical parameters in terms of $n$ or $N$).
\\
This is a very important result, because it shows that the maximum number of information $I$ storable for unit of area is finite and does not depend on the physical parameters of the system. Therefore, it is a property of the spacetime, which emerges from the Heisenberg's uncertainty principle and the equipartition energy theorem in the Planck metric.
In this context, a BH, or more generally any spherical region of spacetime which contains an arbitrary quantity of energy $E$, can not excess $\rho^{EO}_I$ (the ratio $S/A$ remains constant). Thereby, the information stored within an isolated system will be proportional to the surface of a black hole of the same mass.
\\
In this sense, a black hole corresponds to a full energetic system, which saturates the information storable for unit of spacetime on its event horizon.
\\
Finally, let us note that the information stored in a certain surface is a propriety of the system and completely defines its status, like its mass or energy. Naturally, whenever a change in the energy of the system takes place, we observe a proportional variation in the area of the EO and therefore in its entropy and information content, in order to maintain the ratio $S_{BH}/A_{BH}$ constant.
\\
Of course, the information density achieves its maximum value (saturates) on the EO of the BH. Anyway, we can always calculate its value at any equipotential surfaces, by the simple equation:
\begin{eqnarray}
&& \rho_I(d)   \,=\, \frac{1}{8\pi} \frac{M \nabla^2 \Phi(d) }{T_g}   =\frac{n}{d} \frac{2G K_u^2}{T_p}     \,\, \label{eq:rhod}  \, ,
\end{eqnarray}
 where $d$ is the distance from the center of mass of the system.\\
The information density is therefore proportional to the linear degrees of freedom of the system and decreases linearly from the center of mass of the system. Naturally, when $d=R_S \,=\, (2nl_p)$, this expression reduces to Eq. (\ref{eq:SA}).
\\
In this context, it is interesting to observe that Eq. (\ref{eq:rhod}) is directly correlated with the corresponding gravitational field ($\nabla \Phi(d) \,=\, GM/d^2$) of such equipotential surface and therefore defines a dynamics.

Now, following the idea and the model proposed by Verlinde, it is possible to show that there are strong indications that gravity and spacetime are emergent properties in a holographic description of the universe \cite{holo1,holo2,verlinde1,verlinde2,HoloHooft,string,Fea,Feb}.
In particular, whenever an entropy gradient is present in the space, an entropic force is generated, in order to redistribute the matter-energy content present and maximize the total entropy. We interpret this redistribution process as the gravitational force acting between the systems \cite{verlinde1,Fea,Feb}.
\\
In this framework, the entropic force is an effective macroscopic force that originates in a system with many degrees of freedom by the statistical tendency to increase its entropy and does not depend by fundamental fields or by the microscopic dynamics.
\\
Verlinde has found this relation to be equal to \cite{verlinde1}:
\begin{eqnarray}
&&F_{e}= T \cdot \nabla S     \,\, \label{eq:Fe}  \, .
\end{eqnarray}
This means that, in order to have a nonzero force, we need to have a non vanishing temperature. In particular,  whenever an entropy gradient is present and therefore changes in the information content are expected, an entropic-gravitational force arises (is dynamical generated).
\\
At this regards, let us note that Eq. (\ref{eq:Fe}), is generally valid near to the event horizon of the black hole, where a particle of mass $m$ approaches the holographic screen and is subjected to the entropic force generated by the product of the Unruh temperature and $\nabla S$.
Following this prescription, we can generalize this concept to any system placed at any distance $d$ from the event horizon of the other one.
If the systems are not black holes, we can always calculate the corresponding Schwarzschild radius and apply the Verlinde definition to it.
\\
In particular, let us consider two systems $A$ and $B$. The entropic force acting between them can be calculated as the product of the gradient $\nabla S_A$, calculated on its event horizon, and the corresponding Unruh temperature generated by $B$ and measured in $A$ and viceversa.
\\
In other words, we take into consideration the mutual interaction between the systems $A$ and $B$ as the interaction between the corresponding entropy and temperature experimented at the surface of the corresponding event horizon: $F_e=\nabla S_A T_{g_B}$ or $F_e=\nabla S_B T_{g_A}$, where $T_{g (A,B)}$ corresponds to the modified Unruh temperature of Eq. (\ref{eq:Tg}).
This interaction can be expressed in terms of linear degrees of freedom $n_{1,2}$, as follows:
\begin{eqnarray}
&&F_{e}=  \nabla S_{1} T_{g_2} \,=\  n_{1}n_{2} \frac{\alpha_e}{d^2}           \,\, \label{eq:Fe2}  \, ,
\end{eqnarray}
where $S_1$ and $T_{g_2}$ are respectively given by Eq.s (\ref{eq:Sbh}) and (\ref{eq:Tg}), $n_1$ and $n_2$ are the linear degrees of freedom of the systems, $d$ is the distance between them and $\alpha_e= (E_p \cdot l_p) \,=\, 3.16 \cdot 10^{-26}$ (J$\cdot$m) is a constant. At this regards, let's note that the ratio $\alpha_e/\hbar$ returns the speed of the light, therefore $\hbar$ and $\alpha_e$ can be considered as a measure of the energy per unit of spacetime in the Planck metric. Moreover, note that the product $n_1 \cdot n_2$ has the dimension of bits and corresponds to the information stored in the two systems, weighted for the respective linear degrees of freedom $n_1$ and $n_2$.
\\
It is easy to check that Eq. (\ref{eq:Fe2}) converges to the newtonian gravity in the non-relativistic limit ($F_g=GMm/d^2$). However, this formalism is also consistent with a full-relativistic description of the system, as shown by Verlinde \cite{verlinde1,entropy1}.

Finally, let we underline some interesting properties which emerge from this holographic description.
\\
In particular, in the aforementioned hypothesis and in the non-relativistic limit ($F_e\equiv F_g$), we can express the entropy gradient as: $\nabla S_2 =F_g/T_{g_1} = M_2 \cdot g_1/T_{g_1}$, where $g_1$ and $T_{g_1}$ are respectively the gravitational field and the modified Unruh temperature of the first system on the second one.
\\
At this regards, let us observe that the ratio between the temperature and the gravitational field, measured on an arbitrary equipotential surface, is constant for any Schwarzschild black hole.  This constitutes a very important aspect of the model, in fact, being $g/T_g= \alpha_T$, with $\alpha_T=2\pi c^2/T_p l_p \simeq\, 2.5 \cdot 10^{20}$ (m/sec$^2 \cdot$K), the entropy gradient can be simply written as: $\nabla S = \alpha_T M\,=\, n \cdot m_p\alpha_T$ and the entropy of the system as $S=( \alpha_T M) R_S$, which corresponds to Eq. (\ref{eq:Sbh}).
The mass $M$ and the entropy gradient $\nabla S$ are therefore proportional to each other through the simple equation $M= \nabla S/\alpha_T$.
Hence, any variation in the energy spectrum of the black hole, determines a variation in the entropy of the system, which generates an entropic force which tends to redistribute the matter-energy content present, in order to maximize the total entropy. In this scheme, we interpret the mass as a measure of the energetic cost of this redistributing process.
In particular, every time we have an entropy gradient in the system, an equivalent mass (energy) is generated.
In presence of massless particles, like photons, we can always associate to it an equivalent gravitational mass of $m_{\gamma}= E_{\gamma}/c^2$.
\\
Following this scheme, the surface information density on the event horizon can also be expressed as: $\rho^{EO}_I= \alpha_T K_u /4\pi$ which reduces to $ \rho^{EO}_I= n^2 k_b/ R^2_S$, in agreement with Eq. (\ref{eq:SA}).

\section{Conclusion and discussion}
\label{sec:conclusion}

In this work we have proposed a semiclassical analysis of spacetime, based on the simple discretization of its metric in Planck units.
This allows us to simplify many aspects of the theory and study the properties of the black holes in a very simple and elegant way.
\\
In this context, the Schwarzschild metric emerges in a beautiful and natural way, only requiring the  saturation of the energy  at the maximum value compatible with the Heisenberg's uncertainty principle in the Planck metric.
\\
This allows us to introduce and reinterpret many kinds of physical parameters of the system from a new point of view. In particular, black holes become very simple objects, well described in terms of linear degrees of freedom $n$.
\\
One of the key points of this work is the use of the equipartition energy theorem in order to determine the energy distribution content of the system on its event horizon (holographic principle) \cite{holo1,holo2,verlinde1,verlinde2,HoloHooft,string}. It is precisely the possibility to describe the system with a smaller number of dimensions ($D-1$) that simplifies the discussion and makes it possible to obtain, in the hypothesis of quasi-static, spherical and thermodynamical equilibrium, the main physical parameters and equations of general relativity without resorting to complex models.
\\
In this context, we have interpreted the Unruh's radiation as a measure of the vacuum energy of that region of spacetime. This energy is assumed to be generated by quantum fluctuations, induced by the Heisenberg's uncertainty principle, in the spacetime curvature.
\\
In particular, whenever the curvature is negligible (empty space), the average vacuum energy can be neglected (is approximately zero), due to the almost complete annihilation between particles and anti-particles pairs.
\\
Contrariwise, when the curvature increases, the energy (information) content for unit of spacetime becomes larger, until it reaches the maximum value possible compatible with the Heisenberg's uncertainty principle.
\\
In this sense, spacetime can be considered as a "dynamic medium of energy (information) storage". In particular, in agreement with general relativity, the curvature is a manifestation of the energy content of that region of spacetime. When the energy increases up to the Planck level, the curvature becomes large enough to form a BH.

Finally, in the last part of the work, we have presented a holographic description of the system.\\
In this framework, we have introduced the concept of surface information density, through which we have demonstrated the existence of a maximum number of information storable for unit of area. We have shown that  this is a propriety of the spacetime and a constant for any Schwarzschild BH.
In particular, whenever the spacetime achieves $\rho^{EO}_I $, a BH is formed. In this context, a black hole saturates the information and the energy density of spacetime at the maximum level possible, compatible with the Heisenberg's uncertainty principle and the equipartition of the energy in the Planck metric.
\\
Furthermore, let us note that, in a holographic description, the information stored in the BH might not be lost during the gravitational collapse, but, as remarked, it may be stored on the event horizon in form of bits of information  \cite{evap1,evap2}.
This fact opens important implications, not only for holographic models, and a more in-depth analysis is required.
\\
Lastly, we have generalized the concept of entropic force to any system which interacts with another one.
In particular, following \cite{verlinde1}, we have interpreted  this energy redistribution process, provided by Newton's potential, as the gravitational force acting in a non-relativistic situation which tends to maximize the entropy content of the system.
In this situation, we have shown that mass and entropy gradients are related by a relationship of direct proportionality. In particular, every time an entropy gradient takes place, a mass-energy is generated. Therefore, we have interpreted the mass as a measure of the energetic cost of this redistributing process.

Although many aspects still remain to be clarified, we believe this simple model can be considered as a good starting point and gives interesting insights for more complete and exhaustive future analysis.

\end{document}